\documentclass[11pt]{elsart}
\usepackage{epsfig}
\newlength{\capwidth}
\newcommand{\gv}{$(GeV/c)^{2}$}
\newcommand{\abst}{$ \mid t \mid $}

\newcommand{\sbar}{$\overline{s} \ $}

\newcommand{\pt}{$p_{t} $\ }
\newcommand{\ssbar}{$s-\overline{s} \ $}
\newcommand{\bbbar}{$b-\overline{b} \ $}

\newcommand{\lb}{$\Lambda_{b}$ }
\newcommand{\lam}{$\Lambda$ }
\begin{document}
\begin{frontmatter}
\title{Hyperon Polarization in a Quark-Quark Scattering Model}
%\author{Homer A. Neal and Eduard de la Cruz Burelo\\
%{\em Department of Physics, The University of Michigan, Ann Arbor, Michigan 48109}} 
\author{Homer A. Neal\corauthref{cor1}} %and 
\author{and Eduard de la Cruz Burelo\corauthref{cor2}}
\corauth[cor1]{haneal@umich.edu}
\corauth[cor2]{Eduard.Burelo@cern.ch} 
\address{Department of Physics, The University of
Michigan, Ann Arbor, Michigan 48109}
%{\it{ February 6, 2006}}

\begin{abstract}

A continuing mystery in particle physics is the large polarization observed in inclusive hyperon production in high energy proton-proton 
collisions. Both the inclusive nature of the process and the high energies involved make the observation of huge spin effects very 
surprising. In this paper the authors advance the concept that the polarization effects can be accounted for in a simple model based on 
quarks becoming polarized in the direct q-q scattering process and where, at a fundamental level,  the s quark is treated no differently 
than the u and d, except that it is required to have been created as part of a $s-\bar{s}$ pair prior to the collision.
The q-q polarization parameters are extracted from p-p elastic scattering using a previous
model and these are used to  predict the \lam polarization from the p-p data. Remarkably good agreement
is obtained, raising the possibility that the large hyperon polarizations are no
more of a mystery than the p-p polarization itself and that, at least at small \pt, the two may be simply related. This concept is extended in an attempt to explain the polarizations observed in other inclusive hyperon processes with the finding that hyperon and elastic polarizations, which previously seemed almost unrelated, may have a simple and common origin. 
\end{abstract}
\end{frontmatter}

%\maketitle

\section{Introduction}

It was first reported in 1978 that $\Lambda $ \  hyperons inclusively produced in the reaction $p + Be \rightarrow  \Lambda + X $ \ at several hundred GeV/c incident proton momenta exhibited a polarization which grew in magnitude with $p_t$,\  and reached values  near -30\% at \pt  = 1 (GeV/c)~\cite{bunce}. What made this result so surprising was that the $\Lambda$ \  spins aligned so strongly, even at such large incident proton energies, and did so even though the $\Lambda$s were inclusively produced, where all final spin states are summed. Explanation of these measurements, and of spin effects in similar hyperon studies, posed a considerable challenge for all relevant theoretical models. Indeed, to this date, no model has satisfactorily explained these observations, or even successfully related the measurments within the hyperon system itself.

At the \pt  values relevant to the present analysis, it is commonly assumed that the u and d quarks in the outgoing $\Lambda $ \  derive from the incident proton. Further, the s-quark in the outgoing \lam is believed to originate from a \  \ssbar pair produced in the vacuum during the hadronization process. This picture, augmented by the fact that SU(6) requires the (u,d) pair to be in a singlet state ~\cite{lach}, is taken in many models to imply that whatever spin effects there are in the inclusive \lam production, they are due to some special feature of the s-quark. Several models have been created using this paradigm \cite{soffer}. Many are successful in accounting for one or more aspects of the polarization, but fail in explaining significant other features.

In a series of papers Neal and Nielsen (e.g., \cite{neal1},~\cite{neal2}) \ have argued that, on the basis of an analysis of elastic p-p polarization measurements, one can identify the fundamental polarization parameter for quark-quark scattering for the light quarks. Indeed, these authors suggest that the fractal pattern observed in the p-p data is simply a manifestation of multiple q-q scattering. The constraints imposed by such a pattern make it possible within the context of this model to extract a presumed q-q polarization with considerable sensitivity. 

The present paper raises the question of whether the observed $\Lambda $ polarization seen in reactions in the several hundred GeV domain can be predicted by using the p-p polarization at similar energies. We claim that the results are sufficiently intriguing to suggest a careful review of the assumptions of previous models for $\Lambda$ \ inclusive polarization. We also suggest that an analysis of inclusive \lb polarizations at the Tevatron and LHC may lead to valuable insights into the actual subprocesses involved in heavy quark production and decay. 

The approach we take in the model advanced here is to assume that to produce a \lam with a given \pt (\ $< 1 GeV/c$ \ ) there must be a collision involving the s-quark from a just-created \ssbar pair where sufficient transverse momentum is given to it to provide both the resulting \lam and the \sbar with their required transverse momenta, assuming the \ssbar have essentially zero relative momentum until the ultimate hadronization takes place. We then look at the analog p-p elastic scattering process and note what the polarization would be for the equivalent q-q scattering inside the proton would be. We then scale the corresponding polarization to apply to the s quark using the model of multiple scattering of Szwed~\cite{Szwed}. 

The first test of this concept would be to compare the observed \lam polarization with the predicted polarization using the p-p data. Very good agreement is obtained for both the shape and magnitude of the \lam polarization. This has encouraged us to extend the concept to address other unresolved features of the hyperon system.

\section {Special Features of Inclusive Hyperon Polarization}

Hyperon production is clearly quite a complex process which encompasses how quark-antiquark pairs are produced, how three quarks interact, what role flavor plays, how quarks within a baryon scatter, and how they hadronize. Spin could play a role in each of these subprocesses,in just a few, or potentially in none. One is inclined to dismiss the last option, however, since such huge spin effects are observed experimentally, so the quest really is to determine from what source these effects derive. No simple picture has emerged after decades of study that can explain many of the large effects that have been observed. 

We summarize below several of the salient observations when hyperons $H $ \  are inclusively produced in $p + Nucleus \rightarrow H+X $ reactions in the multi-hundred GeV region.

\begin{figure}[htb]                                                            
\begin{center}                                                               
\mbox{\epsfig{figure=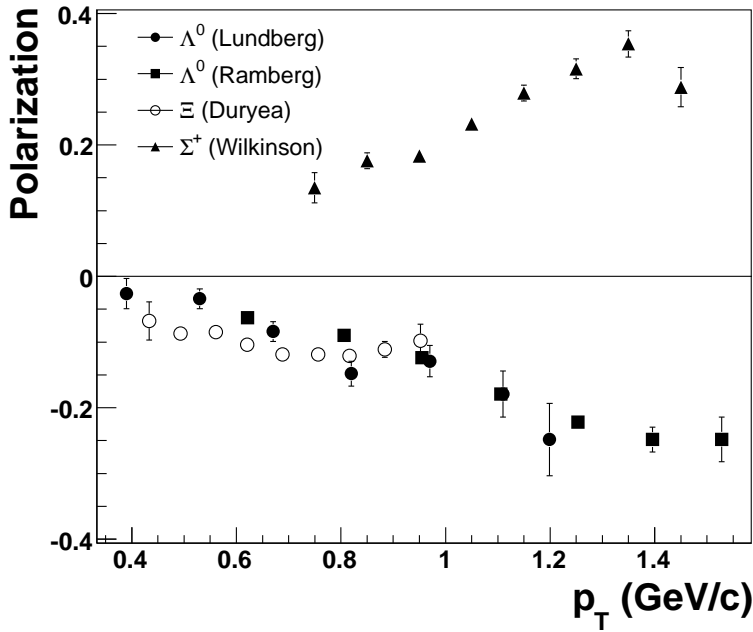,height=3.5in}}                               
\parbox{12cm}{                                                            
\caption{Illustration of the similarity of the \lam (Ref.~\cite{lundberg} and \cite{ramberg}) and $\Xi $ \ polarization~\cite{duryea}, 
and the different structure of the $ \Sigma$  polarization~\cite{wilkinson}.}                                                                  
\label{ejaht1a}                                                                
}                                                                              
\end{center}                                                                  
\end{figure}

\begin{itemize}

\item inclusive $\Lambda$  polarization at a given \pt has little dependence on the initial proton energy.

\item inclusive $\Lambda$  polarization has little dependence on the target material.

\item the polarization of inclusive $\Lambda$ s is negative and grows more negative linearly with \pt until a plateau is reached near \pt = 1 (GeV/c).

\item the plateau in \lam polarization for constant x extends from \pt=1 (GeV/c) to the highest measured \pt

\item the polarization of $ \Sigma $ hyperons is positive in the region of 1 (GeV/c), opposite to the situation for $\Lambda$ 
hyperons (see Figure~\ref{ejaht1a}).

\item the polarization of the $\Xi$ has a clear similarity to that of the $\Lambda$ in spite of its very different quark constituent 
structure (see Figure~\ref{ejaht1a}). 

\end{itemize}

Hyperons are distinguished by which three quarks they contain, and the specific total spin J of a subset of two of the three quarks (i.e., diquarks). Thus, hyperon systems could provide a fertile testing ground for single, two and three quark systems, providing that a reliable analysis framework is developed.

\begin{figure}[htb]                                                            
\begin{center}                                                               
%\mbox{\epsfig{figure=HyperonPol.eps,height=3.0in}}                               
%\mbox{\epsfig{figure=HyperonPol.eps,height=6.0in,width=3.2in}}                               
\mbox{\epsfig{figure=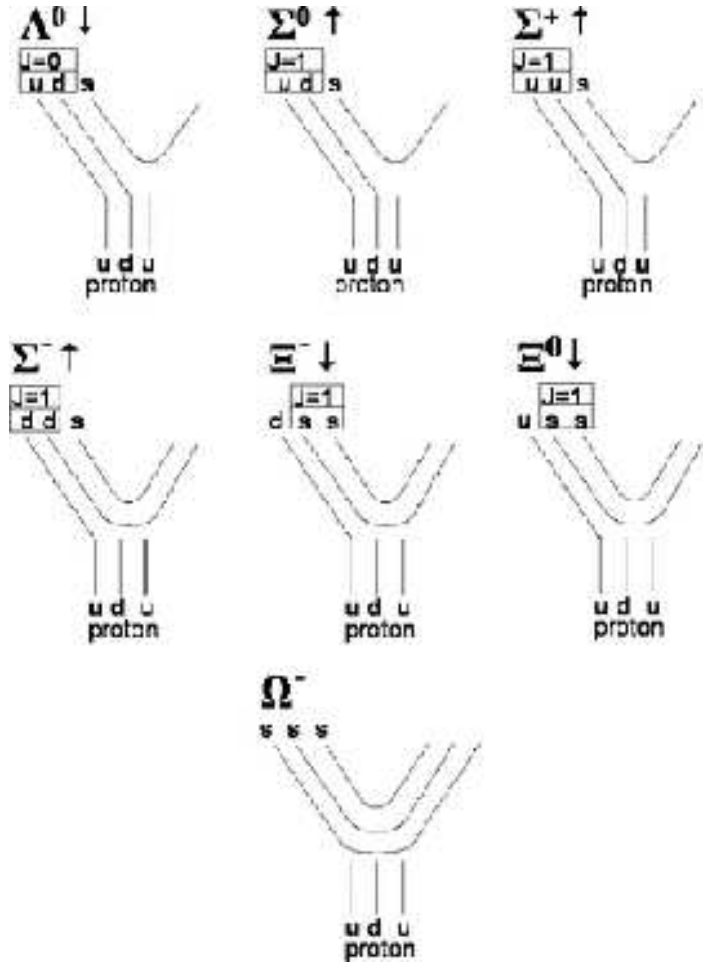,height=5.0in}}                               
\parbox{12cm}{                                                            
\caption{Hyperon Polarization Diagrams (adapted from Lach~\cite{lach}).
\label{qcontent}}
                                                  
}                                                                              
\end{center}                                                                  
\end{figure}

In the \pt region $ < 1 (GeV/c) $  (called hereafter Region I) one can roughly characterize inclusive polarizations as being either negative or positive. Those that are negative, for example, seem to not just be negative, but almost identically negative. Figure~\ref{qcontent} summarizes the quark content and polarization behavior and illustrates the challenge of seeking a simple explanation for the spin behavior of the different members of the hyperon family. In the interest of searching for any relevant systematics, it is interesting to note that:

\begin{itemize}

\item all cases having one s quark correspond to positive polarizations, except for the \lam which is the only instance when a single s quark is arrayed against a J=0 diquark state

\item all cases with two s quarks correspond to negative polarization ( and to J=1 diquarks)

\item no correlation exists between the final polarization and the number of beam quark candidates in the final hyperon

%\item all J=1 diquark systems with one s quark correspond to positive polarizations

%\item all J=1 diquark systems with two s quarks correspond to negative polarizations

\item no differences are observed in terms of polarization sign when u and d quarks are interchanged

\end{itemize}

\bigskip

Any successful model should either explain or incoprorate this set of observations. 

\section{Evidence for the dominance of quark-quark scattering in p-p elastic scattering}

In ~\cite{neal1},~\cite{neal2} a model has been advanced which suggests an approach for extracting the polarization in q-q scattering. The authors interpret the structure in the elastic p-p polarization to be suggestive of multiple scattering of constituent quarks where, in a region spanning \abst values up to $\sim 1$ \gv \  it is assumed that one quark scatters from one quark, and in the region (\#2), between $ \sim 1$ \gv\ and $\sim 3$ \gv, two quarks scatter from two. Support for such a claim comes from the dramatic changes of slope in the fixed-angle cross section near $-t = 1$ \gv \ , and the success in reproducing the fractal behavior of the polarization with the functional form that relates the two types of scattering.

\begin{figure}[htb]                                                            
\begin{center}                                                               
\mbox{\epsfig{figure=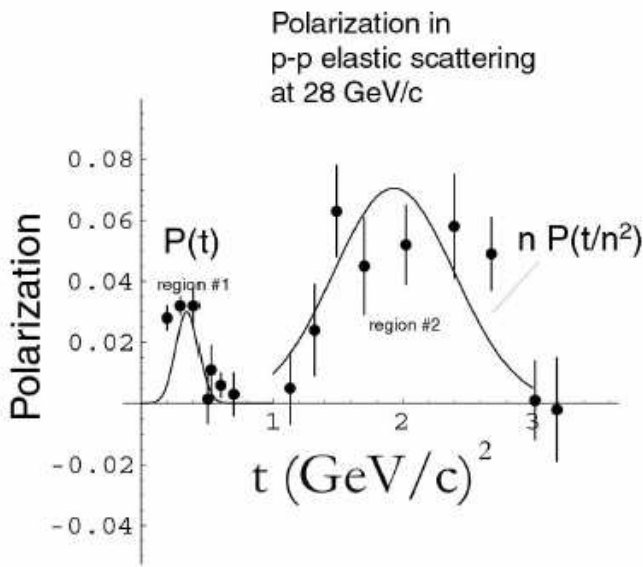,height=3.0in}}                               
\parbox{12cm}{                                                            
\caption{Illustration of the predictive power of the q-q scattering model of ~\cite{neal1} and ~\cite{neal2}  for
the polarization in p-p elastic scattering at 24 and 28 GeV/c ~\cite{pol28}. In the model the polarization
in region \#2 is fully determined by the polarization in region \#1 when the average
number of multiple scatters in region \#2 is specified (n = 2.35).}                                                                  
\label{ejaht1}                                                                
}                                                                              
\end{center}                                                                  
\end{figure}

An example of the predictive power of the model is given in Figure ~\ref{ejaht1}, where the p-p polarization at 24 and 28  GeV/c is shown in the two regions (Ref.~\cite{pol28}). The model requires that if the polarization in region \#1 has the form $P(t)$, the polarization in region \#2 should be of the form $n P(t/n^2)$, where $n$ represents the average number of multiple scatters in region \#2. The fit shown uses $n=2.35$ which, given that contributions are expected from adjoining regions, is suggestive of double scattering. It is the success of such a simple picture in accounting for the structure in p-p elastic scattering polarizations from very low energies (e.g., 6 GeV/c in ~\cite{neal1}) all the way up to the highest energies where measurements have been made that encompass the two \pt regions, that we are motivated to consider the implications for the large spin effects in the inclusive production of \lam hyperons.

\bigskip

\section{ \lam Cross Section; Evidence for Mechanism Change near \pt = 1 (GeV/c) } 

\begin{figure}[htb]                                                            
\begin{center}                                                               
\mbox{\epsfig{figure=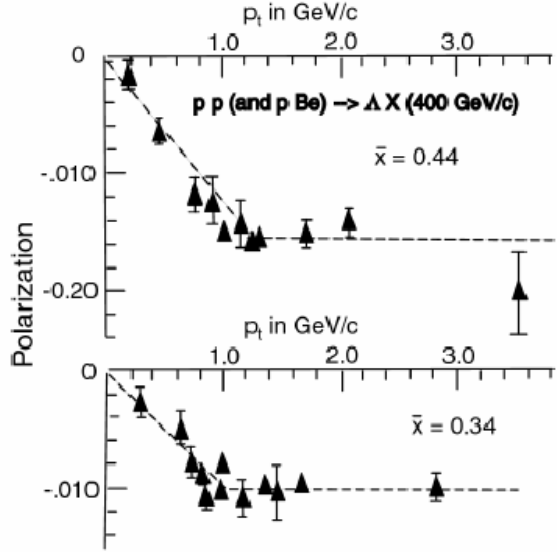,height=3.0in}}                               
\parbox{12cm}{                                                            
\caption{ Illustration of plateau in \lam inclusive polarization at large \pt and at
different values of x. Adapted from Ref. ~\cite{heller}.
}                                                                  
\label{ejaht4}                                                                
}                                                                              
\end{center}                                                                  
\end{figure}

\begin{figure}[htb]                                                            
\begin{center}                                                               
\mbox{\epsfig{figure=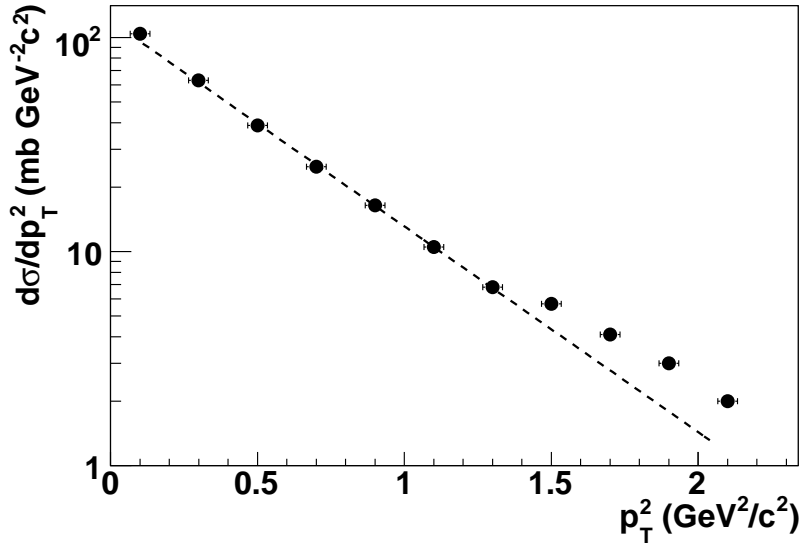,width=4.5 in}}                               
\parbox{12cm}{                                                            
\caption{ Differential cross section of \lam production by neutrons in Copper~\cite{lbcrosssection} as function of $p_{T}^{2}$. 
}                                                                  
\label{ejaht2a}                                                                
}                                                                              
\end{center}                                                                  
\end{figure}

As mentioned in a previous section, 
%and will be reviewed in more detail below, 
it is clear from the \lam polarization data at fixed x that something rather dramatic occurs near \pt $ = 1 (GeV/c) $ (see Figure~\ref{ejaht4}). If, as seems apparent,  a whole new scattering mechanism sets in around this value one would certainly expect to see an effect in the scattering cross section. Surprisingly, very little \lam cross section data exists for \pt values larger than 1 (GeV/c). But, as seen from Figure~\ref{ejaht2a},
where \lam are inclusively produced by neutrons in Copper,
an effect is clearly indicated. It is intriguing to ask whether a change of the scattering method occurs at this \pt because some other mechanism becomes more probable (e.g., such as it becoming easier for two initial state quarks to scatter than one), or if the \lam polarization saturates. Our analysis below  is confined to \pt values in Region I, because of the likelihood that a single model explanation might pertain to this region. This does not mean, of course, that further study might not reveal a relationship between this region and subsequent ones, as was found in the case of p-p elastic scattering discussed above.

\section {Proposed Model Relating p-p Elastic Scattering and \lam Inclusive Production}

In this paper we propose that several of the observations above about \lam production can be explained in terms of scattering of the quarks in the incoming proton. The following concepts form the essence of our model.

\begin{itemize}

\item the primary mechanism for the production of hyperons at low \pt is the scattering of the initial state quarks, whether those quarks are valence quarks or sea quarks.

\item the strange quarks required to produce the individual hyperons are generated via \ssbar pair production within the incident proton  prior to the collision

\item the polarization of the final hyperon at low \pt is the result of the scattering of a previously existing s-quark from the incident proton and its devolution into the final baryon with spin alignment adjustments required to accomodate the proper final diquark spin structure

\item the non-scattered incident quarks which are pulled into the final hyperon do not add to the hyperon polarization

\item the universal light quark polarization parameter can be extracted from the analysis of p-p polarizations, using the approach of Neal and Nielsen (~\cite{neal1}, \cite{neal2})

\item the polarization of the s quark at a given \pt is obtained by the scaling of the light quark polarization by quark mass as in Ref.~\cite{Szwed}.

\end{itemize}

\begin{figure}[htb]                                                            
\begin{center}                                                               
\mbox{\epsfig{figure=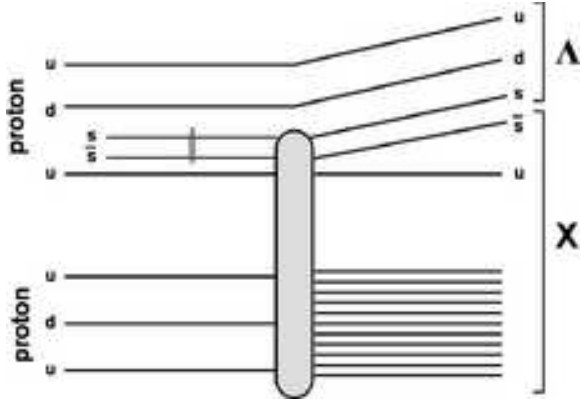,width=3.0 in}}                               
\parbox{12cm}{                                                            
\caption{ Model \lam production mechanism. }                                                                  
\label{MechanismL}                                                                
}                                                                              
\end{center}                                                                  
\end{figure}

One of the first issues to be resolved before one can compare polarizations from p-p scattering to those from inclusive \lam production is the determination of what proton momentum is to be compared to what \lam momentum. Clearly, a simple assumption that one should just compare data at the same \pt could cause one to compare very different internal quark-quark sub-processes. The model assumptions we employed is that the actual production of a \lam occurs as shown in Figure~\ref{MechanismL}. In Region I in p-p scattering we assume, as in \cite{neal1},~\cite{neal2}, that one quark scatters off one quark and the entire \pt of the proton is generated in that scattering and the entire polarization of the proton occurs as a result of that scattering. So the \pt and the polarization of the proton derives from the single scatter. The corresponding \lam production is assumed to occur in a single scattering of a s-quark situated in the incident proton, with that s quark being bound to a J=0 diquark either in the proton or during hadronization. When the s-quark is hit, it acquires a \pt which then provides the \pt of the resulting \lam and of the \sbar quark, which will later disappear as part of some final hadronization process in which it participates. This argument leads to the following relation:
\begin{equation}
k = p_{t\Lambda} + p_{t sbar}\nonumber
\end{equation}

where k is the transverse momentum of the s-quark scatter.

If one assume that a strange quark constituent mass accounts for $\sim 1/2$ the mass of the \lam then one has
\begin{equation}
k = \sim \frac{3}{2}p_{t\Lambda}\nonumber
\end{equation}
 
Operationally, this implies that to compare the proton and \lam polarizations at a certain \lam momentum, one should examine the p-p 
polarization at a momentum that is 1/2 times larger than the \pt of the \lam. Results of such a comparison will be given below, 
following a discussion of the scaling of the polarization itself.

If it is assumed that the scattering process at the momentum k is precisely the same in the p-p and \lam cases, then one would have only to scale the momenta as suggested above and then make a comparison. But, indeed, another model choice must be made. If, in parallel with the elastic p-p case, we assume that only one q-q scatter occurs in the region of interest, which quark is it that scatters, the u, d, or s? As noted above, we have chosen the s-quark. One reason is its presumed larger physical cross section, and the presumed relative ease with which u and d quarks can be acquired during the hadronization process, compared to the comparable rate at which \ssbar pairs of the proper \pt could be produced during hadronization. To obtain the polarization of the s-quark we assume that one must scale the light quark polarization by the ratio of the constitutent masses for the u and s, as suggested by the model of of Szwed~\cite{Szwed}. We note that, with precision data and if the scattering mechanism proposed herein were separately validated, the proposed analysis could directly address the question of how quark-quark polarization depends on quark mass.  

We will assume in the analysis below that the polarization of the s-quark, and thus of the \lam, at the momentum transfer k is given by 

\begin{equation}
P(k) = \frac{m_s}{m_u}P_{proton}(k)\nonumber
\end{equation}

This sets the stage for using the p-p polarization to "predict" the \lam polarization. The results are presented in the next section.

\section {Comparison of p-p and Inclusive \lam Polarization Data: Model Results}

\begin{figure}[htb]                                                            
\begin{center}                                                               
%\mbox{\epsfig{figure=lambda56.eps,width=3.0in}}                               
\mbox{\epsfig{figure=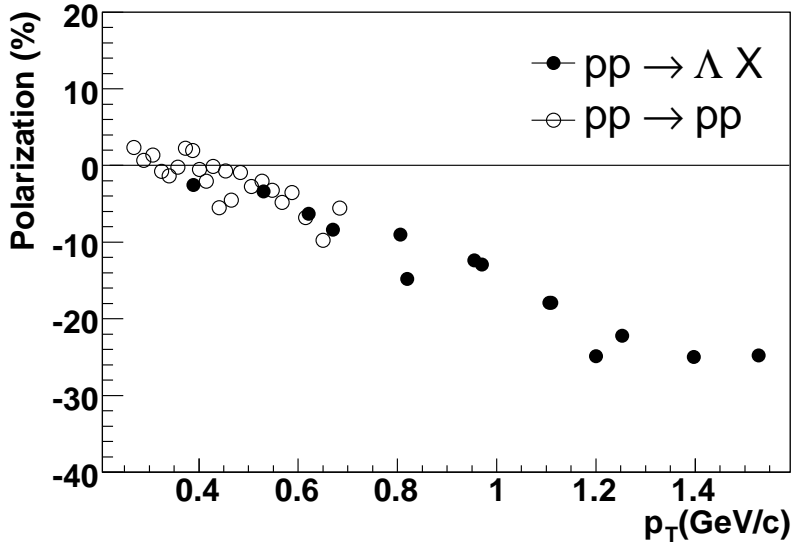,width=4.5in}}                               
\parbox{12cm}{                                                            
\caption{ Lambda polarization measured in \cite{lundberg} and \cite{ramberg}, and the values of the \lam polarization predicted
by the described model using as input the p-p elastic polarization data of ~\cite{synder}.
}                                                                  
\label{ejaht2}                                                                
}                                                                              
\end{center}                                                                  
\end{figure}

A critical test of the above model assumptions is to check whether the \lam inclusive polarizations imputed from p-p polarizations fits the actual \lam polarization data at the corresponding \pt. We conduct this test by assuming that the s -quark is scattered by an incoming quark and thereby becomes polarized. As discussed above, we assume that in this \pt Region I it is easier for a single quark scattering to take place than a scattering from the diquark or from individual u, or d quarks. We superimpose on actual \lam polarization data the prediction of what the \lam polarization should be as imputed from each p-p polarization data point in the measurements of ~\cite{synder} at 300 GeV/c.  The premise of the basic scattering process is provided in (\cite{neal1},~\cite{neal2}). Thus, in making the comparison in Figure~\ref{ejaht2}, we have applied the factor 2/3 to the \pt of the p-p data before plotting. Also, before plotting, the measured p-p polarization value is multiplied by a factor of  2 to account for the fact that the constituent mass of the s-quark is roughly twice that of the light quarks. 

Figure~\ref{ejaht2} shows a remarkable congruence of the predicted and measured  values of the \lam polarization in the region where data exist for both the p-p and \lam \ reactions. Moreover, the fact that the slopes of the data are independently similar adds further support for the idea that the underlying processes are the same. A priori, there is no reason to expect that the transformed p-p data would align in this way with the measured \lam polarization data. This suggests that, at least in the \pt range explored, the \lam polarization may be due simply to the independent polarization acquired by the s quark in its quark-quark scattering.

\section {Explanation of the Plateau in the Inclusive $\Lambda$ Polarization}

\begin{figure}[htb]                                                            
\begin{center}                                                               
\mbox{\epsfig{figure=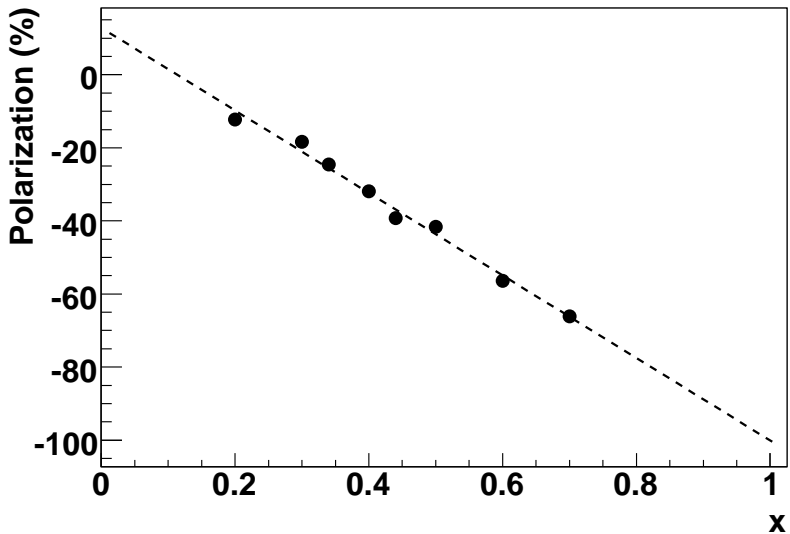,height=3.0in}}                               
\parbox{12cm}{                                                            
\caption{Projection of \lam polarization saturation as function of $x$, assuming a dilution of 41\%. Data points are extracted from Figure 5 in~\cite{heller}.}                                                                  
\label{ProjPol}                                                                
}                                                                              
\end{center}                                                                  
\end{figure}

Another striking feature of the \lam polarization is that, though it varies rapidly with \pt for values less than $\sim $ 1 GeV/c, it achieves a plateau near that \pt value and the plateau is maintained all the way to \pt values near 3.5 GeV/c (see Figure~\ref{ejaht4}) \cite{heller}. This behavior persists for different values of the Feynman variable x, the ratio of the \lam momentum to the proton beam momentum. In the spirit of the model, we ascribe this behavior to a saturation of the spin of the s-quark at 100\% for the nearly elastic ( x = 1) component of its scatters. Of course, the \lam polarization does not itself reach 100\% because at x values less than 1.0 many other processes presumably take place with no associated spin preferences. Since certain models (e.g.,~\cite{Szwed}, discussed above)  suggest that polarizations at the q-q level should scale with the quark mass, it is not unrealistic to assume that 100\% polarization of the s-quark could occur near \pt$\sim$1.2 GeV/c, close to where the saturation is observed to set in for the \lam polarization. A simple linear extrapolation of the plateau value of the polarization vs. x shows that a \lam polarization of 100\%  (see Figure~\ref{ProjPol}) might occur for x = 1 at a \pt value near 1.2 GeV/c if we assume a polarization dilution value of 41\%. It is interesting to note that 30\% is the value reported in Gourlay et al\cite{gourlay} as the contribution to the \lam sample from the "non-prompt" process $ \Sigma^0 \rightarrow \Lambda \gamma $, which is thought to be the principal such "non-prompt" process , and Siebert et al reported a \lam sample contamination of 50\% from the same source~\cite{siebert}. The possibility that s-quark becomes 100\% polarized through prompt processes near \pt$\sim$1.2 GeV/c must be taken seriously.
 
%(Figure on s-quark polarization projection)
%\begin{figure}[htb]                                                            
%\begin{center}                                                               
%\mbox{\epsfig{figure=plateau55.eps,height=3.0in}}                               
%\parbox{12cm}{                                                            
%\caption{ Illustration of plateau in \lam inclusive polarization at large \pt and at
%different values of x. Adapted from Ref. ~\cite{heller}.
%}                                                                  
%\label{ejaht4}                                                                
%}                                                                              
%\end{center}                                                                  
%\end{figure}

We note that a large s-quark polarization in Region I would drive  not only the large polarization seen in \lam production, but in $\Xi$ \ and $\Sigma $ as well, assuming that a single s-quark scattering also occurs in their production (even though the $\Sigma $ polarization is positive, as discussed in Section 9). 

One model that provides an interesting suggestion for the origin of the plateau is that of the Lund model~\cite{Lund} by Andersson et 
al. 
In that model \lam\'s are produced as the u and (ud) quarks in the incoming proton are separated by the collision, causing a color force tube to snap and produce an \ssbar pair. Field theory requires that the $s$ and $\bar{s}$ not be produced at the same point. Moreover, it is assumed that the s quark would preferentially have its \pt roughly aligned with the \pt of the outgoing \lam. Since the \pt of the s and \sbar must be equal and opposite, an angular momentum is created, which must then be cancelled by the spins of the \ssbar pair. The model assumes that the spins of the s and \sbar are preferentially  parallel and that their polarization grows until 100\% is achieved. The plateau thus emerges naturally in the model, with saturation of the \lam polarization occurring around \pt of $\sim 1 (GeV/c)$. The model has other attractive features, such as possibly providing an explanation for why the polarization seems to be so insensitive to the momenta of the colliding protons. Indeed, if it is the color tube breaking process that gives rise to the \ssbar pair, then there is no real reason why the polarization \pt dependence should depend strongly on the incident proton momenta. However, in recent years one major criticism have been raised against this model. 
%One [REF] is that its premises can not be cast in relativistic terms. The other is the 
Results from Jefferson Lab experiments~\cite{carman} suggest that $q-\bar{q}$ pairs are produced with the quark and anti-quark spins 
anti-aligned, not 
aligned as required by the Lund model. These apparent defects leave us with essentially no current model that can effectively explain this dramatic feature of inclusive \lam production. The model we advance in this paper raises the question as to whether a color tube breaking process is really needed to explain the data, since it appears that the \lam processes are not really that different from those of p-p elastic scattering, if the s-quark exists just before scattering in the proton.

\section {Similarity of the $\Lambda$ and $\Xi$ Polarizations}

An additional longstanding mystery in inclusive hyperon production has to do with the dramatic similarity of the polarization of the 
\lam and $\Xi$ hyperons. Their quark structure is quite different, with the \lam being $ s-(ud)_0 $ and the $\Xi^0$ \ being $ u-(ss)_1 $. The suggestion from our model that the s quark becomes highly polarized in Region I offers a plausible explation for the \lam - $\Xi$ paradox. If the (ss) diquark has a large z-component of polarization because at least one of the s quarks was highly polarized from the scatter, that would induce a $\Xi$ \ polarization similar to that seen in the \lam case. That the extent of the negative polarization of the $\Xi$ so closely tracks that of the \lam suggests that the spins of the three quarks involved have the same basic origins -- one s quark scatter while the others observe.

\section {Why is the $\Sigma $ Polarization Positive?}

Another in the list of mysteries is why the polarization of the $\Sigma $ hyperon is large and positive, while its sister \lam is large 
and negative (see Fig.~\ref{ejaht1a}). Structurally speaking, the only difference between the two is that in one case (\lam) the diquark 
is a 
$(ud)_0$ \ and in the other ($\Sigma $ ) a $(ud)_1$. Both the $\Sigma $ and \lam have a single unassociated s quark. As has been pointed 
out by previous authors~\cite{wilkinson2} this feature can be explained through the angular momentum structure of a 1/2 x 1 system. 
Specifically, a 
strongly negative polarized s quark will selectively enhance the importance of the state where both the u and d in the J=1 diquark have their spins aligned in the positive direction, giving the overal $\Sigma $ a positive polarization. The new feature we are suggesting is, again, that the s quark is highly polarized through the scattering process and that the gap in the $\Sigma $ and \lam polarization values is a natural result of this process.  There is only one s-quark scatter in each case. 
\section {The  $\Omega$ \ Polarization}

The $\Omega$ \  consists of three valence s-quarks.  Given that its spin is 3/2, there are no known symmetries that would prevent the three quarks from having all spins aligned and thus resulting in a large polarization for the $\Omega$ \, if indeed the s-quark is  the carrier of  significant polarization in the production process.

\begin{figure}
\begin{center}                                                               
\mbox{\epsfig{figure=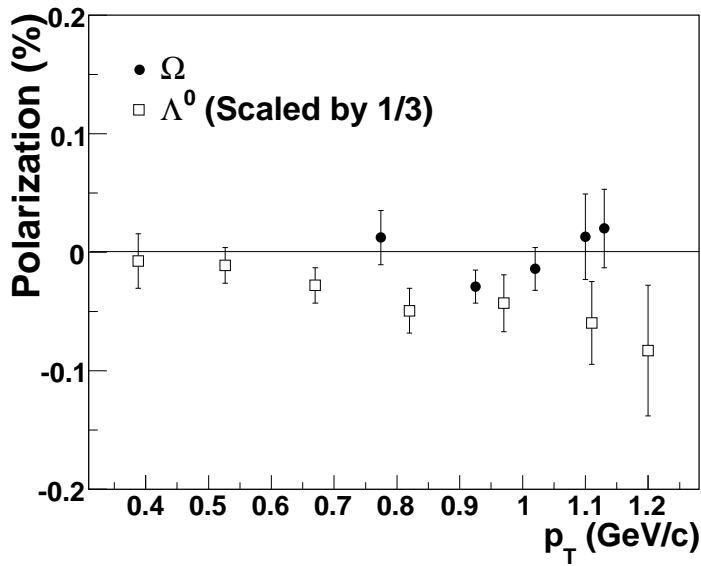,width=4.0in}}                               
\parbox{12cm}{                                                            
\caption{ $\Omega$ polarization in function of \pt. \lam polarization scaled by $1/3$ is shown on top of the $\Omega$ polarization.}
\label{omega:lambda}                                                                
}                                                                              
\end{center}         
\end{figure}

A mystery that has been long discussed is that, in spite of the expectation that large polarizations might be anticipated from highly polarized s-quarks, as suggested by the \lam data, the $\Omega$ \  seems to have a polarization consistent with zero, as shown in Fig~\ref{omega:lambda}.

In the model we have advanced herein this small value of the $\Omega$ \ polarization follows somewhat naturally. We assume that, as we do for all of the hyperon processes in Region I, there is only the scattering of one quark and this is the only quark that can become directly polarized. This immediately sets an upper bound of 33\%  on the polarization for the $\Omega$. An interesting test is to superimpose the lambda polarization divided by 3 on the $\Omega$ \  polarization, since one saturated s-quark could polarize the \lam at 100\%, but only 33\% for the $\Omega$. As one sees in Figure~\ref{omega:lambda}, the match is quite good. For the $\Omega$ \ to be 100\% polarized all three of its s quarks would need to point in the same transverse direction. In the context of our model, only one s quark scatters and can thus get polarized. This ceiling on the maximum value, coupled with dilutions of the polarization from non-prompt processes, leads to predicted values of the $\Omega$ \ polarization that compare favorably with observation.

\section {The Vanishing $\overline{\Lambda}$ Polarization }

In the early days of the observation of hyperon polarization, the fact that large polarizations were seen in inclusive \lam production 
and that no polarization was observed in $\overline{\Lambda}$ production was taken to mean that only valence quarks can be polarized. 
This, of course, is now known not to be the case.  Since our model explicitly admits the scattering of non-valence quarks as a mechanism 
for producing polarized hyperons, it is proper to inquire how the  zero  $\overline{\Lambda}$ polarization can be explained  in the 
model. However, it is important to note that nowhere in the model have we attempted to explain the dynamics of quark-quark scattering or 
how polarizations arise in that scattering process. We have only illustrated the commonalities in the different 3-quark systems which we 
assume are all related through common q-q scaterring processes. We make no claim that q-q and q-$\bar{q}$ processes are simply related. 
Indeed, the former could give rise to large polarizations, while the latter could give polarizations which are vanishingly small. The processes are different and models like ours can not be expected to relate the two without the introduction of detailed dynamics, which is not our purpose here.

\section {Critical Model Tests}

The model we have described can be further tested in a variety of ways. For example, measuring the polarization of the $\Xi \ $, $\Sigma \ $, and $\Omega \ $ in both Regions I and at \pt values greater than ~1.0 (GeV/c) would be extremely valuable. Significant discontinuities are expected if indeed the entire scattering mechanism changes near \pt = 1(Gev/c) as we suggest. Also, better data at low \pt for p-p elastic and \lam inclusive could permit the comparison of the light quark and s-quark polarizations and to test the proposals that they should be proportional to the constituent or chiral masses.

\section {Implications for $\Lambda_b$ \ Polarization at LHC Energies }

Part of the theoretical complication in understanding \lam spin effects is due to the fact that the three constitutent quarks have roughly the same mass. No one quark dramatically dominates the process kinematically and the relative scale for perturbative vs. non-perturbative QCD effects is not clear. That is not the case for $\Lambda_b$ , where the b quark is almost an order of magnitude more massive than the other participating quarks.  The large mass of the b has implications for the extent to which \lb would be intrinsically polarized as it is produced from a \bbbar pair.

In the work of \cite{gold} an extensive analysis has been done on spin effects to be expected in quark-quark scattering, quark-gluon scattering, and in gluon fusion processes. One general result of this work is that the more massive the quark the greater the polarization effects should be for a given \pt. If this also applies to the b-quark production  processes, very significant spin effects could be present in the \lb production. We note that a central suggestion from the above analysis is that the s-quark itself seems to already appear at several hundred GeV/c with a prompt polarization of 100\%.

In the LHC ATLAS experiment a sample of over 75,000 \lb should be produced in the first three years of running. Through an analysis of the \lb decays ( $\Lambda_{b} \rightarrow J/\psi \Lambda$, with $J/\psi \rightarrow \mu \mu$ and $\Lambda \rightarrow \pi^{-}p$ \ , one should be able to extract the \lb polarization. The resolution of the polarization determination is estimated to be .016 for this sample \cite{atlas}. With such a capability at the LHC one could hope to determine the extent to which polarization might be generated in gluon splitting process as opposed to the scattering processes discussed in the current paper.

Regarding the latter, one should note that we can not be sure that the gluon-splitting processes believed to dominate heavy quark production are not also responsible for polarization effects in light quark scattering. Thus the indications that polarization effects might increase with the mass of the quark makes the b-quark studies at the LHC highly interesting.

\section{Concluding Remarks}

By drawing upon previous work in which the q-q polarization parameter can purportedly be extracted from elastic p-p polarization data, the possibility of explaining some of the prevailing mysteries of inclusive \lam production has been examined. It is found that the slope and magnitude of the \lam polarization predicted from the extracted p-p polarization is very consistent with the \lam polarization data. Moreover, a plausible explanation has been provided for the remarkable plateau effect seen in inclusive \lam data. Furthermore, we find that our presumption that the non-valence quarks participate in the scattering processes on an equal footing with the valence quarks allows the explanation of several spin effects within the entire octet.

The model presented is patently semi-classical. Motivation for attempting such an analysis on a system that is clearly quantum mechanical derives from the remarkable successes in prior studies that treat quarks independently in explaining hadron cross sections, in the analysis of the fractal nature of p-p polarizations that seem to suggest that quark spin effects in a hadron can be treated semi-classically, and that, indeed, macroscopic classical measurements were what led to the discovery of proton spin in the first place.

\section{Acknowledgements}

We wish to thank M. Longo, R. Akhoury, and A.D. Krisch for their advice on various aspects of the above analysis. We also want to thank summer students Robin Smith, Sarah Lumpkins and Jacob Bourjaily for their assistance with this study.

We wish also to acknowledge the support of the U. S. Department of Energy. Finally, we wish to acknowldege the hospitality of the European Organization for Nuclear Research, where much of this work was pursued as we prepare for upcoming LHC spin studies.

\end{document}